\documentclass[12pt]{article}
\pdfoutput=1
\usepackage{draft}

\newtheorem{conjecture}{Conjecture}

\begin{document}

\begin{center}

\hfill \\
\hfill \\
\vskip 1cm

\title{Holomorphic CFTs and topological modular forms
}

\author{Ying-Hsuan Lin$^{a}$ and Du Pei$^b$}

\address{${}^a$Jefferson Physical Laboratory, Harvard University, Cambridge, MA 02138, USA}

\address{${}^b$Center of Mathematical Sciences and Applications, Harvard University, Cambridge, MA 02138, USA}

\email{yhlin@fas.harvard.edu, dpei@cmsa.fas.harvard.edu}

\end{center}


\begin{abstract}
We use the theory of topological modular forms to constrain bosonic holomorphic CFTs, which can be viewed as $(0,1)$ SCFTs with trivial right-moving supersymmetric sector.
A conjecture by Segal, Stolz and Teichner requires the constant term of the partition function to be divisible by specific integers determined by the central charge.
We verify this constraint in large classes of physical examples, and rule out the existence of an infinite set of extremal CFTs, including those with central charges $c=48, 72, 96$ and $120$.
\end{abstract}

\section{Introduction}

Based on earlier work of Segal \cite{segal1987elliptic,Segal}, a conjecture by Stolz and Teichner  \cite{ST1,ST2} states that the space of 2d $(0,1)$ supersymmetric quantum field theories (SQFT) has the structure of an $E_\infty$-spectrum, giving a generalized cohomology theory known as ``TMF'' (see e.g.\ \cite{douglas2014topological} for a comprehensive review). This relation has several interesting consequences for physics described below.

\begin{enumerate}
    \item There are new invariants of 2d $(0,1)$ SQFTs that are ``topological modular forms''. The ring of topological modular forms, denoted by $\pi_*(\text{TMF})$, refines the elliptic genus by a collection of torsion-valued novel invariants.  They are ``complete invariants" in the sense that any two SQFTs with the same TMF invariants can be ``smoothly'' deformed into each other.  
    \item For each topological modular form, there is at least one associated 2d $(0,1)$ SQFT.
    \item More familiar invariants such as the Witten index and the elliptic genus are subject to nontrivial constraints, and many values of these invariants, though legitimate at first sight, \emph{cannot} be realized by a physical SQFT.
\end{enumerate}
By now, various aspects of these consequences have been studied in the literature. For example, regarding the first consequence, the physical meaning of a $\mathbb{Z}_{24}$-valued invariant has been understood \cite{GJW,GJ2}. And the new invariants of 2d $(0,1)$ SQFTs have been used to construct new invariants of smooth 4-manifolds \cite{GPPV}. The second consequence mentioned above predicts the existence of many 2d SQFTs with intriguing properties (see e.g.\ \cite{GJ} for a family of new theories constructed using self-dual ternary codes, and \cite[Sec.~3]{GPPV} for a ``dictionary'' between topological modular forms of small degrees and 2d $(0,1)$ SQFTs).

The main objective of this note is to explore the third consequence when applied to purely bosonic holomorphic conformal field theories (CFT). 
A bosonic holomorphic CFT can be regarded as a $(0,1)$ superconformal field theory (SCFT) with trivial right-moving supersymmetric sector, and the elliptic genus of the SCFT is just the partition function of the bosonic holomorphic CFT.\footnote{By contrast, the earlier work by \cite{GJ} studied right-moving holomorphic SCFTs. A holomorphic CFT can also be put in the left-moving sector, in which case it is not required to be supersymmetric.}
The TMF theory dictates that the elliptic genus of any $(0,1)$ SQFT satisfies a nontrivial divisibility constraint, a fact that was recently used to study the vanishing of global anomalies in heterotic string compactifications \cite{Tachikawa:2021mvw,Tachikawa:2021mby}.  In this note, we invoke such constraints to achieve the following.
\begin{enumerate}
    \item We explain the divisibility of the constant term observed in the partition functions of lattice CFTs, $c=24$ CFTs, and lattice orbifold CFTs.
    \item We argue that many of the extremal partition functions written down by \cite{hoehn2007selbstduale,hohn2008conformal,W} cannot be physically realized.
\end{enumerate}

The rest of this note is organized as follows.  In Section~\ref{sec:divisibility}, we present the key divisibility constraint coming from the TMF theory, and apply it to bosonic holomorphic CFTs.  In Section~\ref{sec:physical}, we verify the divisibility constraint in a number of physical examples.  In Section~\ref{sec:extremal}, we examine the realizability of extremal partition functions.  Finally, Section~\ref{sec:remarks} provides some additional remarks.

\section{Divisibility constraint via TMF theory}
\label{sec:divisibility}

The Stolz--Teichner conjecture associating 2d $(0,1)$ SQFTs with TMF---a theory of topological modular forms---has a myriad of physical consequences. Although the TMF theory may be abstruse to most physicists, here we only need one particular implication that can be described in a physics-friendly manner, as follows.  Denote the gravitational anomaly of the theory by $\nu\in \mathbb{Z}$, which equals $2(c_R - c_L)$ for CFTs.  Under what is called the Witten genus homomorphism, 
\ie\label{WittenGenus}
\{\text{Topological Modular Forms} \}\longrightarrow \{\text{Modular Forms}\},
\fe
the TMF invariant of a theory maps to $\eta(q)^\nu Z_\text{ell}(q)$, where $\eta(q)$ is the Dedekind function, and
\ie 
    Z_\text{ell}(q) = \tr (-1)^F q^{\frac{H-P}{2}} = \tr (-1)^{F_R} q^{L_0 - \frac{c_L}{24}}
\fe
is the elliptic genus (the second expression is valid in SCFT; $H$ is the Hamiltonian, $P$ the momentum, and $L_0$ the 0th Virasoro generator). The Witten genus homomorphism  is a map of graded rings, under which the grading on the left given by $\nu$ becomes twice the weight of the modular form on the right.\footnote{The extra $\eta(q)^\nu$ factor is to cancel the gravitational anomaly, which can be thought of as coupling to free chiral fermions. After including it, the function becomes a modular form in the standard mathematical sense. 
More precisely, it belongs to the ring of weakly holomorphic modular forms (i.e.\ can have a finite number of negative powers of $q$) with integral coefficients. 
} What is interesting about this map is that it is \emph{neither surjective nor injective}, which means that (1) the elliptic genus can be refined, and (2) not every integral modular form can be realized as the elliptic genus of some SQFT.  

Property (2) can be stated more precisely as follows.
The image of the Witten genus homomorphism has an integral basis given by \cite[Proposition 4.6]{hopkins2002algebraic}
\ie\label{basis}
    a_{i,j,k} c_4^i c_6^j \Delta^k, \quad i\ge0; ~j=0,1,
\fe
where $c_4$ and $c_6$ are the Eisenstein series,\footnote{In our convention, they are normalized to have integral coefficients starting with 1.} $\Delta = \eta^{24}$ is the modular discriminant, and
\ie\label{coefficients}
    a_{i,j,k} = 
    \begin{cases}
        \displaystyle \frac{24}{\text{gcd}(24,k)} & \text{if } i=j=0,
        \\
        2 & \text{if } j=1,
        \\
        1 & \text{otherwise}.
    \end{cases}
\fe
We call the requirement that certain coefficients 
in the elliptic genus 
must be divisible by \eqref{coefficients} the \emph{divisibility constraint}.

\subsection{Constraints on holomorphic CFTs}

A bosonic holomorphic CFT with central charge $c = 24n$, $n \in \bZ$, can be regarded as  a $(0,1)$ SCFT with $(c_L, c_R) = (24n,0)$,\footnote{Recall that one can tautologically regard a bosonic theory as a fermionic theory, as the former can be defined on spin manifolds. In fancier language, one can ``fermionize'' the bosonic theory with respect to a trivial $\bZ_2$ global symmetry.} and hence gravitational anomaly $\nu = -48n$.  Holomorphy and modular invariance require the elliptic genus, which is the same as the bosonic partition function, to be a degree-$n$ polynomial in the Klein $j$-invariant
\ie
j(q) = \frac{c_4^3}{\Delta} = q^{-1} + 744 + 196884q + \cO(q^2)
\fe
with integer coefficients,
\ie
Z(q) = \sum_{p=0}^n b_p j(q)^p, \quad b_p \in \mathbb{Z}.
\fe
The Witten genus is then
\ie\label{wg}
\eta^{-48n} Z(q) = \Delta^{-2n} Z(q) = \sum_{p=0}^n b_p c_4^{3p} \Delta^{-p-2n}.
\fe
By comparing \eqref{wg} with the general form \eqref{basis}, we find that the constraint \eqref{coefficients} requires $b_0$ to be divisible by\footnote{
    The divisibility constraint has to be consistent with taking products of theories because $\pi_*$(TMF) is a ring.  In the case of two bosonic holomorphic CFTs with $c = 24m$ and $c = 24n$, this boils down to a basic property of gcds
    \ie 
        \frac{24}{\text{gcd}(24,m+n)} ~\bigg|~ \frac{24}{\text{gcd}(24,m)} \frac{24}{\text{gcd}(24,n)}
    \fe
    for arbitrary $m$ and $n$.  First note that the above is obviously true when $\text{gcd}(24,m,n)=1$, as the product on the right is divisible by 24.  When $\text{gcd}(24,m,n)=d>1$, the above still holds as $d \mid \text{gcd}(24,m+n)$.
}
\ie\label{divisibility}
    \frac{24}{\text{gcd}(24,2n)} =
    \begin{cases}
        1 & n \equiv 0 \pmod {12},
        \\
        12 & n \equiv \pm1 \pmod {12},
        \\
        6 & n \equiv \pm2 \pmod {12},
        \\
        4 & n \equiv \pm3 \pmod {12},
        \\
        3 & n \equiv \pm4 \pmod {12},
        \\
        12 & n \equiv \pm5 \pmod {12},
        \\
        2 & n \equiv 6 \pmod {12}.
    \end{cases}
\fe

\section{Lattice CFTs, Schellekens' list, and lattice orbifolds}
\label{sec:physical}

In this section, we verify the divisibility constraint \eqref{divisibility} in a number of physical bosonic holomorphic CFTs, including lattice CFTs, all $c=24$ theories classified by Schellekens \cite{Schellekens:1992db}, and various $c=48, 72$ lattice orbifold CFTs constructed by Gem\"unden and Keller \cite{Gemunden:2018mkh,Gemunden:2019dtr}.

\subsection{Lattice CFTs}
\label{sec:lattice}

A consistent free chiral boson CFT can be formulated for any given even unimodular lattice.  It is a theorem of Borcherds \cite[Theorem 12.1]{borcherds1995automorphic} that if the lattice dimension is divisible by 24, i.e.\ $24 \mid c$, then the constant term in the $q$-expansion of its partition function is divisible by 24.  In particular, if we consider $n$ copies of the Leech lattice CFT, then according to the theorem, the constant term in the $q$-expansion of $Z_\text{Leech}(q) = (j(q) - 720)^n$ is divisible by 24, which implies that the constant term in the $q$-expansion of $j(q)^n$ is divisible by 24.
Hence, for a general holomorphic CFT (need not be a lattice or lattice orbifold CFT), the divisibility constraint \eqref{divisibility} is the same on the constant term in the $j(q)$ polynomial and on that in the $q$-expansion.  To put the above comments in concrete formula, we define
\ie\label{b}
    Z(q) = \sum_{p=0}^n b'_p q^{-p} + O(q) = \sum_{p=0}^n b''_p J(q)^p,
\fe
where $J(q) = j(q) - 744$.  Then since $b_0 \equiv b'_0 \equiv b''_0 \pmod {24}$, the divisibility constraint \eqref{divisibility} on $b_0$, $b'_0$, and $b''_0$ are all equivalent.  This in turn means that Borcherds' theorem trivializes the divisibility constraint \eqref{divisibility} for lattice CFTs (without orbifolds).

\subsection{Schellekens' list at $c=24$}

The $c=24$ holomorphic CFTs were famously classified by Schellekens \cite{Schellekens:1992db}.  In his table, the possible values of $(\cN =) \, b''_0 = b_0 - 744$ are
\ie 
    0, 24, 36, 48, 60, 72, 84, 96, 108, 120, 132, 144, 156, 168, 192, 216, \\ 240, 264, 288, 300, 312, 336, 360, 384, 408, 456, 552, 624, 744, 1128,
\fe
which are all divisible by 12, consistent with and saturating the divisibility constraint \eqref{divisibility}.

\subsection{Orbifold CFTs by lattice automorphisms at $c=48,72$}

Gem\"unden and Keller \cite{Gemunden:2018mkh,Gemunden:2019dtr} constructed many $c=48$ and $c=72$ holomorphic CFTs by taking asymmetric orbifolds of extremal lattices by lattice automorphisms.
Scanning through their list, we find that $12 \mid b'_0$ for $c = 48$ and $8 \mid b'_0$ for $c = 72$, both consistent with but not saturating the divisibility constraint \eqref{divisibility} (off by a factor of 2 in both cases).

\section{Extremal CFTs}
\label{sec:extremal}

An extremal partition function $Z_n(q)$ at central charge $c = 24n$ was defined by \cite{hoehn2007selbstduale,hohn2008conformal,W} to be the unique polynomial in $J(q)$ such that
\ie
    Z_n(q) = \frac{q^{-n+\frac{1}{24}}}{\eta(q)} (1-q) + \cO(q) = \frac{q^{-n}}{\prod_{i=2}^\infty (1-q^i)} + \cO(q),
\fe
i.e.\ only containing Virasoro descendants up to weight $n$.
The constant term $b'_0$ in the $q$-expansion in $Z_n(q)$ is given by (recall the definition of $b'_0$ from \eqref{b})
\ie\label{pp}
    b'_0 = p(n) - p(n-1),
\fe
where $p(n)$ is the ``partition function'' (counting the number of ways to write $n$ as a sum of non-negative integers).\footnote{Note that $p(n) - p(n-1)$ counts the number of partitions of $n$ without 1's.}
The divisibility constraint \eqref{divisibility} applied to the above can be concisely written as (recall from Section~\ref{sec:lattice} that $b_0 \equiv b'_0 \pmod {24}$)
\ie 
    \frac{24}{\text{gcd}(24,2n)} ~\bigg|~ p(n) - p(n-1).
\fe
We find that up to $n = 100$, the values of $n$ satisfying this condition are 
\ie
1,6,9,10,12,13,18,22,24,28,36,37,40,48,52,54,60,66,72,75,82,84,90,96,100.
\fe
Interestingly, the first few values after $c=24$, famously realized by the Monster CFT \cite{frenkel1984natural,frenkel1989vertex}, are ruled out, including $c=48, 72, 96, 120$.\footnote{The genus two partition functions of hypothetical $c=48, 72$ extremal CFTs were determined in \cite{Gaiotto:2007ts}.  It was proved in \cite{Gaiotto:2008jt} that a hypothetical $c=48$ extremal CFT cannot have Monster symmetry.}

\section{Remarks}
\label{sec:remarks}

By viewing bosonic holomorphic CFTs as $(0,1)$ SCFTs with trivial right-moving sector, we applied the divisibility constraint \eqref{divisibility} coming from the TMF theory to rule out the realizability of the extremal partition functions for a large set of central charges.  However, it is clear that the constraint \eqref{divisibility} has nothing to say about the realizability of the extremal partition functions when $24 \times 12 = 288 \mid c$, and hence it remains possible that an infinite family of extremal holomorphic CFTs with increasing central charge exists, and may be dual to a theory of pure gravity in AdS$_3$.

TMF constraints on bosonic holomorphic CFTs with $24 \mid c$ are the most general in the following sense.\footnote{We thank Theo Johnson-Freyd for a discussion regarding this point.}
First, a bosonic holomorphic CFT with $8 \mid c$ but $24 \nmid c$ is not subject to the divisibility constraint \eqref{coefficients} because $c_4$ must always present and $c_6$ absent, i.e.\ $i>0$ and $j=0$.
Second, a fermionic holomorphic CFT $V$ with $8 \mid c$ can be bosonized to a pair of bosonic holomorphic CFTs $V_+$ and $V_-$, with $Z_\text{ell}^V(q) = Z^{V_+}(q) - Z^{V_-}(q)$, and hence the TMF constraints on $V$ are guaranteed by those on the bosonic $V_+$ and $V_-$.  Finally, any fermionic holomorphic CFT with $8 \nmid c$ has trivial elliptic genus.

One interesting future direction is to generalize the divisibility constraint to 2d CFTs that couple to a non-trivial 3d topological quantum field theory. This would incorporate general rational CFTs and theories with global symmetries. Then instead of a topological modular form with a map to a modular form, there will be a certain (equivariant) TMF-module equipped with a map to a vector-valued modular (or Jacobi) form. We expect that the divisibility constraint will be richer in this more general setting.

By now, various aspects of the conjectured relation between TMF and 2d $(0,1)$ SQFTs have been tested, and many special cases have been understood (see \cite{Theo} for a state-of-the-art review).  We believe that the divisibility constraint lies very close to the heart of this relation, because (1) unlike many TMF phenomena that factorize through KO theory, it is a genuinely two-dimensional phenomenon, and (2) it is closely related to the 576-periodicity of TMF. Furthermore, the divisibility constraint specialized to bosonic holomorphic CFTs, i.e.\ \eqref{divisibility}, can be stated purely mathematically without invoking stable homotopy theory as follows.

\begin{conjecture}
    For every rational vertex operator algebra with central charge $c = 24n$ and only one irreducible module, the constant term in the character is divisible by $\frac{24}{{\rm gcd}(24, 2n)}$.
\end{conjecture}
It is conceivable that the above conjecture can be settled using purely vertex operator algebra techniques. This would lead to a better understanding of the physical reasons behind the divisibility constraint, which would enable us to not only exclude extremal CFTs using fundamentally physical principles, but also gain deeper insights into the beautiful connection between TMF and 2d $(0,1)$ SQFTs.

\section*{Acknowledgements}

We are grateful to Nathan Benjamin, Chi-Ming Chang, Davide Gaiotto, Sergei Gukov, Theo Johnson-Freyd, Yuji Tachikawa, and Cumrun Vafa for comments on the draft.  YL thanks the hospitality of Caltech during the progress of this work.  YL is supported by the Simons Collaboration Grant on the Non-Perturbative Bootstrap.  

\paragraph{Data availability statement}  Data sharing not applicable to this article as no datasets were generated or analyzed during the current study.

\providecommand{\href}[2]{#2}
\providecommand{\doihref}[2]{\href{#1}{#2}}
\providecommand{\arxivfont}{\tt}

\end{document}